\documentclass[showpacs,amsmath,amssymb,twocolumn,floatfix,prl]{revtex4}
\usepackage[dvips]{graphicx}
\input{epsf}

\begin{document}

\title{Frenkel-Kontorova model with cold trapped ions}


\author{I. Garc\'ia-Mata$^{(1)}$, O. V. Zhirov$^{(2)}$ and D. L. Shepelyansky$^{(1)}$}
\homepage[]{http://www.quantware.ups-tlse.fr}
\affiliation{$^{(1)}$Laboratoire de Physique Th\'eorique, UMR 5152 du CNRS, 
Universit\'e  Paul Sabatier, 31062 Toulouse Cedex 4, France\\
$^{(2)}$Budker Institute of Nuclear Physics, 
630090 Novosibirsk, Russia}

\date{June 6, 2006}
%\date{\today}

\begin{abstract}
We study analytically and numerically the properties of one-dimensional
chain of cold ions placed in a periodic potential of optical lattice
and global harmonic potential of a trap. 
In close similarity with the Frenkel-Kontorova model, a transition from
sliding to pinned phase takes place with the increase of
the optical lattice potential for the density of ions
incommensurate with the lattice period. Quantum
fluctuations lead to a quantum phase transition and melting of 
pinned instanton glass phase at large values of dimensional Planck constant.
The obtained results are also relevant for a Wigner crystal placed
in a periodic potential.
\end{abstract}
%32.80.Lg Mechanical effects of light on atoms, molecules, and ions
%32.80.Pj Optical cooling of atoms; trapping
%63.70.+h Statistical mechanics of lattice vibrations and displacive phase transitions
%61.44.Fw Incommensurate crystals

\maketitle

Nowadays experimental techniques allow to store thousands of cold ions 
and observe various ordered structures formed by
Coulomb repulsion in ion traps \cite{walther1992}.
These structures include the one-dimensional (1D) Wigner crystal, 
zig-zag and helical structures in three dimensions.
The Cirac-Zoller proposal 
of quantum computations with cold trapped  ions \cite{cz1995}
generated an enormous experimental progress in this field
with  implementations of quantum algorithms and quantum state
preparation with up to 8 qubits \cite{blatt,wineland}.
In these experiments \cite{blatt,wineland} ions 
form a 1D chain placed in a global harmonic potential
which frequency $\omega$ determines the eigenfrequencies of
chain oscillations being independent of ion charge
\cite{dubin,morigi}. Highly accurate experimental control
of the chain modes allows to perform quantum gates
between internal levels of ions.  
In addition to ion traps, modern  laser techniques allow
to create periodic optical lattices and store
in them thousands of cold atoms (see e.g. \cite{bloch}).
A single ion dynamics in an optical lattice has been also studied
experimentally \cite{walther1997}. The combination
of these two techniques makes possible to study experimentally the
properties of a 1D chain of few tens of  
ions placed in an optical lattice
and a global harmonic potential at ultra low temperatures.

In this Letter we analyze the physical properties of such a system
and show that it is closely related to the Frenkel-Kontorova Model (FKM)
\cite{fk1938} which gives a mathematical description
of various physical phenomena including crystal dislocations, 
commensurate-incommensurate phase transitions,
epitaxial monolayers on a crystal surface, magnetic chains and
fluxon dynamics in Josephson junctions
(see \cite{braun2004} and Refs. therein). As in the classical FKM
the ion chain exhibits the Aubry analyticity breaking
transition \cite{aubry} when the amplitude of optical potential
becomes larger than a certain critical value.
Below the critical point the classical chain can slide
(oscillate) in the incommensurate optical lattice
while above the transition it becomes pinned 
by the lattice and a large gap opens in the spectrum
of phonon excitations. Above the transition the positions of ions
form a devil's staircase corresponding to a fractal Cantor set
which replaces a continuous Kolmogorov-Arnold-Moser (KAM) curve
in the phase space below the transition.
According to \cite{aubry}
the FKM ground state is unique but in the pinned phase
there are exponentially many stable configurations
which are exponentially close to the ground state
and the system resembles a fractal spin glass 
\cite{zhirov2002}. In the quantum FKM \cite{borgonovi,berman,bambihu}
the instanton tunneling between quasidegenerate  configurations
leads to a melting of the pinned instanton glass 
and the quantum phase transition into sliding phonon phase
takes place above a critical value of the effective Planck constant $\hbar$
\cite{zhirov2003}. In contrast to the FKM with
nearest neighbor interactions between particles, the 
ions have long range Coulomb interactions. In spite of that
we show that the ion system is effectively described by the
FKM and has similar  phase transitions which can be studied
experimentally. To analyze the properties
of the FK Ion Model (FKIM) with few hundreds of ions we use 
various  numerical methods including classical and Quantum Monte Carlo (QMC)
algorithms as described in  \cite{borgonovi,zhirov2002,zhirov2003}.

The dimensionless FKIM Hamiltonian has the form:
\begin{equation}
H = \sum_{i=1}^{N} (\frac{P_i^2}{2}+\frac{\omega^2}{2} x_i^2 - K \cos x_i) +
\sum_{i > j}\frac{1}{|x_i-x_j|} 
\label{eq1}
\end{equation}
where $P_i, x_i$ are ion momentum and position, $K$ gives the strength
of optical lattice potential and all $N$ ions are placed in a harmonic
potential with frequency $\omega$. To make a transfer 
from (\ref{eq1}) to dimensional
physical units one should note that the lattice constant $d$ in 
$K \cos (x_i/d)$ is taken to be unity, the energy  $E=H$
is measured in units of ion charge energy $e^2/d$
and $\omega^2 \rightarrow m \omega^2 d^3/e^2$ 
where $m$ is ion mass. In the quantum case
$P_i= -i \hbar \partial/\partial x_i$ with dimensionless $\hbar$ 
measured in units $\hbar \rightarrow \hbar/(e \sqrt{md})$.
In the quantum case we use the approximation of distinguishable
ions which is well justified when the distance between
ions imposed by the harmonic potential is comparable with
the lattice period \cite{cz1995}. 

\begin{figure}[t!]
\epsfxsize=3.2in
\epsffile{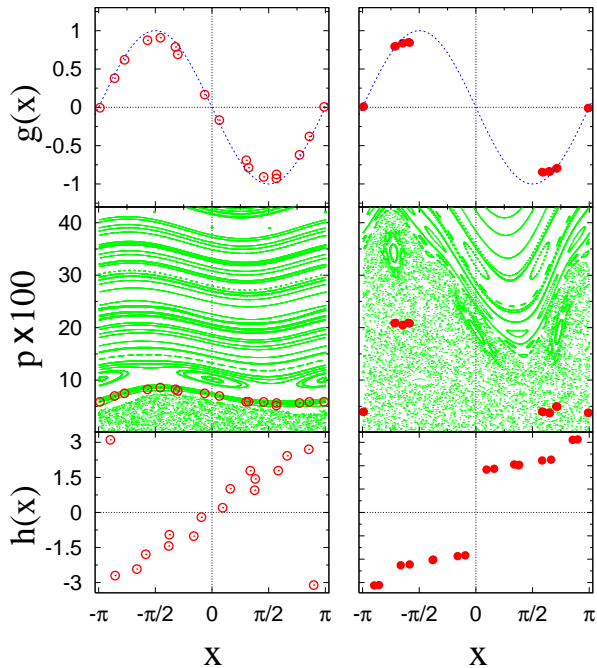}
\vglue -0.0cm
\caption{(Color online) Functions related to the dynamical map (\ref{eq2})
obtained from the ground state equilibrium positions $x_i$ of $N=50$ ions
for $\omega = 0.014$ ($\omega^2=1.986 \times 10^{-4}$)
at $K=0.03$ (open circles, left column) and  $K=0.2$ 
(full circles, right column).
Panels show: the kick $g(x)$ function (top);
the phase space $(p,x)$ of the map (\ref{eq2}) with $g(x)=-\sin x$
(green/gray points) and actual ion positions (red/black circles) (middle);
the hull function $h(x)$ (bottom). The ion positions
are shown as $x=x_i (mod 2\pi)$ for the central 
1/3 part of the chain. 
}
\label{fig1}
\end{figure}
  
We start the discussion from the classical case.
Here the stable configurations
with minimal energy have $P_i=0$ and satisfy the conditions
$\partial H/\partial x_i = 0$.  In approximation of only
nearest neighbor interacting ions these conditions lead
to the dynamical recursive map for equilibrium ion positions $x_i$:
\begin{eqnarray}
p_{i+1} = p_i + K g(x_i) \; , \; \; x_{i+1} = x_i+1/\sqrt{p_{i+1}} \; ,
\label{eq2}
\end{eqnarray}
where the effective momentum conjugated to $x_i$ is
$p_i = 1/(x_{i}-x_{i-1})^2$ and the kick function
$K g(x)=-\omega^2 x - K \sin x$.
To check the validity of this description
we find the ground state configuration using numerical methods
described in \cite{aubry,zhirov2002}. The harmonic
frequency $\omega$ is chosen in such a way that in the middle of the chain
the ion density $\nu = 2\pi/(x_1-x_0)$ at $K=0$
is equal to the golden mean value
$\nu = \nu_g = (\sqrt{5}+1)/2$. This corresponds to an
incommensurate phase with the golden  KAM curve usually studied for the
Aubry transition \cite{aubry,braun2004}. For a fixed $\nu \sim 1$ the strength
of harmonic potential $\omega^2$ drops with $N$ according 
to relations found in  \cite{dubin,morigi}. The numerical values $x_i$
allow to determine  the function $g(x)$ using relations  (\ref{eq2})
and show that with good accuracy $g(x)=-\sin x$ (see Fig.\ref{fig1}). 
We attribute this result to an effective screening of
global harmonic potential 
by long range interactions between ions in the central part
of the chain where the ion density is approximately constant.
For small values of optical potential ($K=0.03$)  
the chain is in the sliding phase with a continuous KAM curve
in the plane $(p,x)$ while above the Aubry transition at $K=0.2$
it is in the pinned phase and forms a fractal Cantor set
in the phase space (Fig.\ref{fig1}). The qualitative change
of chain properties is also seen via the hull function $h(x)$
which gives the ion positions in periodic potential
vs. unperturbed positions at $K=0$ both taken $\mod 2\pi$:
for $K<K_c$ one has a continuous function $h(x) \approx x$
while for $K>K_c$  the hull function has a form of devil's staircase
with clustering of ion positions at certain values
(Fig.\ref{fig1}). 
More  data for a larger number of ions $N=150$ are
shown in Appendix (see Fig.\ref{fig6}).
\begin{figure}[t!]
\epsfxsize=3.2in
\epsffile{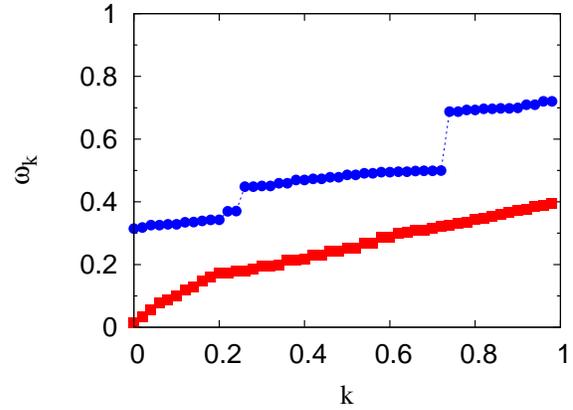}
\vglue -0.0cm
\caption{(Color online) 
Phonon spectrum $\omega(k)$ as a function of
scaled mode number $k=i/N$ ($i=0,\ldots,N-1$),
for $K=0.03$ (bottom curve, squares) and 
$K=0.2$ (top curve, points) for the case of Fig.\ref{fig1}.
}
\label{fig2}       
\end{figure}

\begin{figure}[t!]
\epsfxsize=3.2in
\epsffile{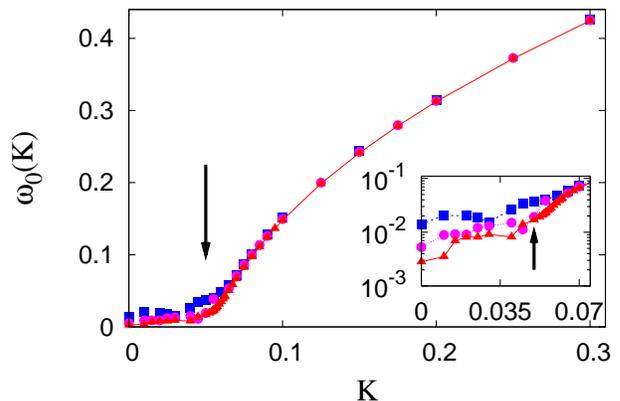}
\vglue -0.0cm
\caption{(Color online) Minimal excitation frequency
$\omega_0(K)$ as a function of periodic potential strength
$K$ for the golden mean ion density $\nu_g$ and 
number of ions $N=50$ (squares; $\omega=0.014,
\omega^2=1.986 \times 10^{-4}$),
$N=150$ (circles; $\omega=0.00528, \omega^2=2.794 \times 10^{-5}$),
$N=300$ (triangles, $\omega=0.00281, \omega^2=7.913 \times 10^{-6}$).
The critical point  $K_c \approx 0.05$ is marked by arrow;
inset shows data near $K_c$.
}
\label{fig3}       
\end{figure}

The structural changes in the chain below and above the Aubry transition
at the critical value $K_c$  are also clearly seen in the spectrum of
phonon excitations shown in Fig.\ref{fig2}
(see also Fig.\ref{fig7} in Appendix for $N=150$). For $K<K_c$
the spectrum of phonon modes has a sound like form at small wave vectors
$k$, going down to a minimal oscillation frequency $\omega_0 \approx \omega$,
which is small compared to frequencies of periodic potential.
For $K>K_c$ the spectrum of excitations is characterized by a 
phonon gap when $\omega_0$ becomes  independent of number of ions $N$
(Fig.\ref{fig3}). According to the data of Fig.\ref{fig3} for 
$\omega_0(K)$ and of Fig.\ref{fig8} in Appendix for the hull function
the critical point is located at $K_c \approx 0.05$.
The value $K_c$ can be obtained approximately by 
a reduction of map (\ref{eq2}) to the Chirikov standard map
\cite{chirikov1979}.  For that the second equation (\ref{eq2})
can be linearized near $1/\sqrt{p} = 2\pi/\nu_g$ 
that gives the Chirikov standard map with chaos
parameter $K_{eff} = K(2\pi/\nu_g)^3/2$ and 
the critical value $K_c = 0.034$ corresponding to $K_{eff} = 1$.
This value is smaller than the value $K_c \approx 0.05$
obtained from numerical data that can be attributed to 
a strong dependence of $K_{eff}$ on $\nu_g$.

\begin{figure}[t!]
\epsfxsize=3.2in
\epsffile{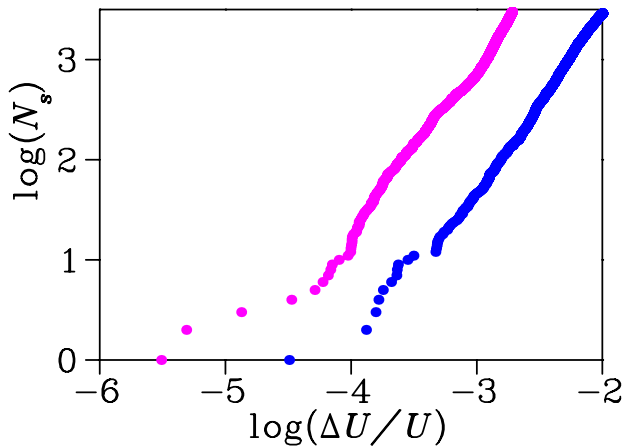}
\vglue -0.0cm
\caption{(Color online) 
Number of equilibrium configurations $N_s$
as a function of their relative excitation
energy $\Delta U/U$ above the ground state
for 50 (blue/black) and 150 (magenta/gray) ions at $K=0.2$
with $\omega$ as in Fig.\ref{fig3}
(logarithms are decimal).
}
\label{fig4}       
\end{figure}

The obtained results show that the ion chain in a periodic potential
has close similarities with the FKM. The map (\ref{eq2}) describes
quite accurately the central 1/3 part of the chain
where the ion density is approximately constant.
Outside of this part the ion density starts to grow
and deviations from the map start to be visible
in $(p,x)$ plain and $h(x)$, even if the $g$-function remains
rather close to $g=-\sin x$. A separation on two parts
is also seen in the lowest phonon eigenmodes which
are localized in the central part of the chain
(see Figs.\ref{fig9},\ref{fig10} of Appendix).
Also, in similarity with the FKM \cite{zhirov2002}, 
we find that in the pinned phase at $K>K_c$ the FKIM 
has properties of a spin glass with enormous number
of stable equilibrium configurations $N_s$
in the very close vicinity of the ground state
(see Fig.\ref{fig4}).

\begin{figure}[t!]
\epsfxsize=3.2in
\epsffile{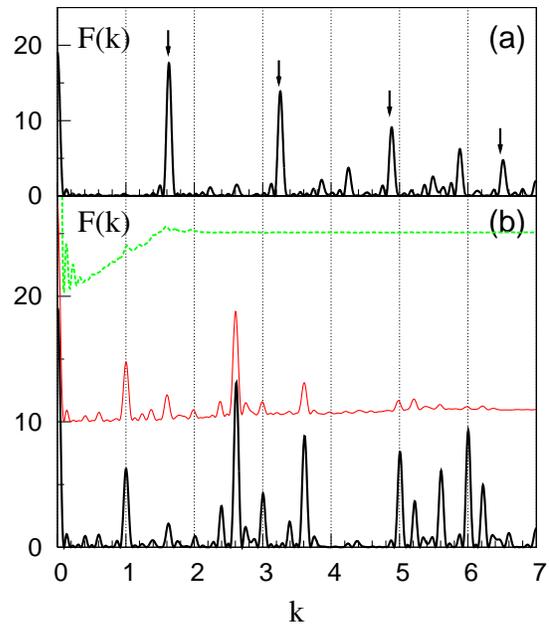}
\vglue -0.0cm
\caption{
(Color online) Formfactor $F(k)$ (see text) of the chain  with
$N=50$ ions and $\omega=0.014, \omega^2=1.986 \times 10^{-4}$.
\textbf{\textsf{(a)}}: The classical incommensurate phase at $K=0.03$, 
$\hbar=0$, arrows mark the peaks at 
integer multiples of golden mean density $\nu_g$.
\textbf{\textsf{(b)}} The  pinned phase case at $K=0.2$ 
for $\hbar=0$ (bottom black curve), $\hbar=0.1$ (middle red curve
shifted 10 units upward), $\hbar=2$ (top green curve shifted 20 units upward,
for clarity $F(k)$ is multiplied by factor 5). 
The temperature of the quantum chain
is $T=\hbar/\tau_0$ with $\tau_0=400$ 
so that $T \ll K$ and $T \ll \hbar \omega_0(K)$.
}
\label{fig5}
\end{figure}

These quasidegenerate configurations
become very important in the quantum chain
where quantum tunneling between them
leads to nontrivial instanton excitations.
To study the quantum case we use the QMC approach
\cite{creutz,borgonovi,zhirov2003}
with the Euclidean time $\tau$ in the interval
$[0, \tau_0]$ and  system temperature
$T=\hbar/\tau_0$. We use large values of 
$\tau_0$ (e.g. $\tau_0 \ge 400$) so that 
the temperature $T$ is small compared to
periodic potential $K$ and phonon gap $\hbar \omega_0$.
To see the structural changes produced by quantum 
fluctuations it is convenient to compute the
formfactor of ion positions defined as 
$F(k)= \langle  \sum_i \big| \exp(i k x_i(\tau) ) \big|^2 \rangle  /\delta N$,
where summation is taken over the central part of the chain
with $\delta N \approx N/3$ ions, $x_i(\tau)$ are 
ions positions at Euclidian time $\tau$ and brackets
mark averaging over $\tau$. In the classical case
$x_i$ are the equilibrium positions of ions. For $\hbar=0$
the formfactor peaks are equidistant in $\nu_g$ that clearly show the
incommensurate sliding phase at $K=0.03 < K_c$ (Fig.\ref{fig5}a).
For $K=0.2 > K_c$ the peaks in $F(k)$ are at integer $k$ values
showing that ions are pinned by the optical lattice (Fig.\ref{fig5}b bottom).
In the quantum case at $K=0.2 > K_c$ and 
small values of $\hbar=0.1$ the density of instantons
is small and quantum tunneling does not destroy the pinned
phase. However, at large $\hbar=2$ quantum fluctuations
lead to melting of pinned phase, the integer peaks disappear
and  $F(k)$ becomes continuous (Fig.\ref{fig5}b). This means that the ion chain can slide 
in the optical lattice. The quantum phase transition 
from pinned instanton glass to sliding phonon gas
takes place at zero temperature and certain $\hbar_c \sim 1$.

The formfactor $F(k)$ can be experimentally obtained by 
light scattering on the ion chain and in this way  various
phases of the system can be detected. For experimental conditions
like in \cite{blatt} the distance between $Ca$ ions is about
$\Delta x \approx 5 \mu m$  so that the golden mean density $\nu_g$
corresponds to the lattice constant 
$d = \nu_g \Delta x/2\pi \approx 1.3 \mu m$.
This value can be realized by laser beams crossed at a fixed
angle. The transition to pinned phase takes place
at the optical lattice potential $V = K_c e^2/d \approx 0.6 K$
that in principle can be reached in strong laser fields.
For ${^{40}}Ca^{+}$ ions with such $d$ the dimensionless Planck constant
is $\hbar_{eff} = \hbar/(e\sqrt{md}) \approx 3 \times 10^{-5}$.
This means that such experiments with cold trapped ions
can be performed in a deep semiclassical regime hardly
accessible to the QMC numerical simulations.
We also note that a strong gap in
phonon spectrum $\omega_k$ for $K>K_c$ (Figs.~\ref{fig2},~\ref{fig3})
may be useful for protection of quantum gate operations
in quantum computations. 

Higher values of dimensionless $\hbar$
can be reached with electrons forming a Wigner crystal
placed in a periodic potential where  $\hbar \approx 0.1$ for
$d \approx 5 nm$. Such a situation may appear in
1D electron wires, molecular structures or nanotubes
where a devil's staircase behavior has been
discussed recently \cite{novikov}. In the regime when
the number of electrons per period is of the order
of $\nu \sim 1$ the effects related to statistics of particles
are not so crucial and the quantum melting of the pinned phase
should qualitatively follow the scenario described here.
Hence, the obtained results describe also
a more general problem of a Wigner crystal in a periodic
potential.

\begin{acknowledgments}
This work was supported in part by the EC IST-FET project EuroSQIP.

\end{acknowledgments}

%\newpage
%\begin{widetext}

%\appendix
%\newpage
\section{APPENDIX}
\begin{figure}[t!]
\epsfxsize=3.2in
\epsffile{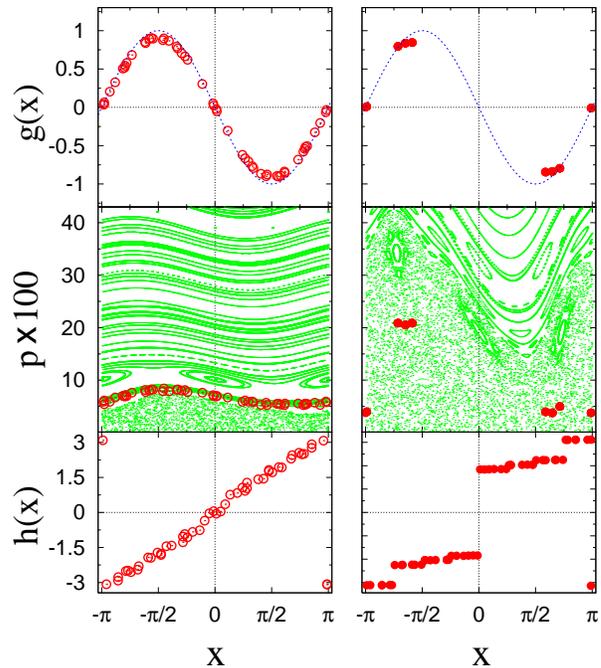}
\vglue -0.0cm
\caption{(Color online) Same as in Fig.\ref{fig1} but for 
$N=150$ ions at  $\omega=0.00528$
($\omega^2=2.794 \times 10^{-5}$)
corresponding to the golden mean
density of ions in the middle of the chain.
}
\label{fig6}
\end{figure}

\begin{figure}[t!]
\epsfxsize=3.2in
\epsffile{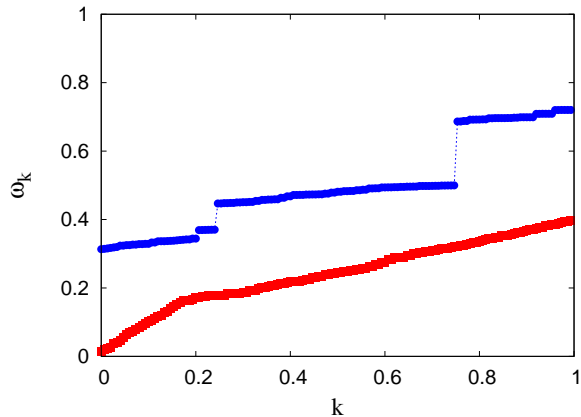}
\vglue -0.0cm
\caption{(Color online) 
Phonon spectrum $\omega(k)$ as a function of
scaled mode number $k=i/N$ ($i=0,\ldots,N-1$),
for $K=0.03$ (bottom curve, squares) and 
$K=0.2$ (top curve, points) for the case of Fig.\ref{fig6}
at $N=150$ (compare with data of Fig.\ref{fig2} for $N=50$).
}
\label{fig7}
\end{figure}

\begin{figure}[t!]
\epsfxsize=3.2in
\epsfysize=3.2in
\epsffile{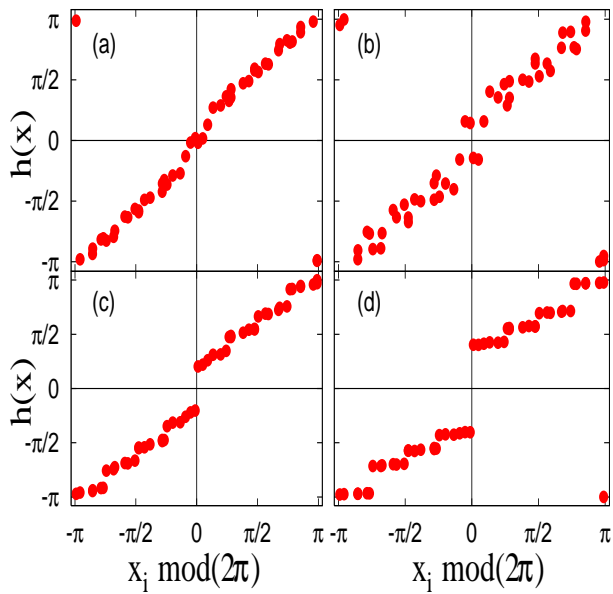}
\vglue -0.0cm
\caption{
(Color online) Evidence of the Aubry transition in the hull function.
The hull function $h(x)$ 
shows the ion positions $x_i(K)$ mod $(2 \pi)$ in presence
of periodic potential as a function of ion positions  
$x_i^{(0)}$ mod$(2 \pi))$ without potential ($K=0$).
Here $N=150,\ \omega=0.000528,  \omega^2=2.794\ 10^{-5}$,
\textbf{\textsf{(a)}} $K=0.04$;
\textbf{\textsf{(b)}} $K=0.055$;
\textbf{\textsf{(c)}} $K=0.065$;
\textbf{\textsf{(d)}} $K=0.1$.
Only the central $1/3$ part of the chain
is shown.
}
\label{fig8}
\end{figure}

\begin{figure}[t!]
\epsfxsize=3.2in
\epsffile{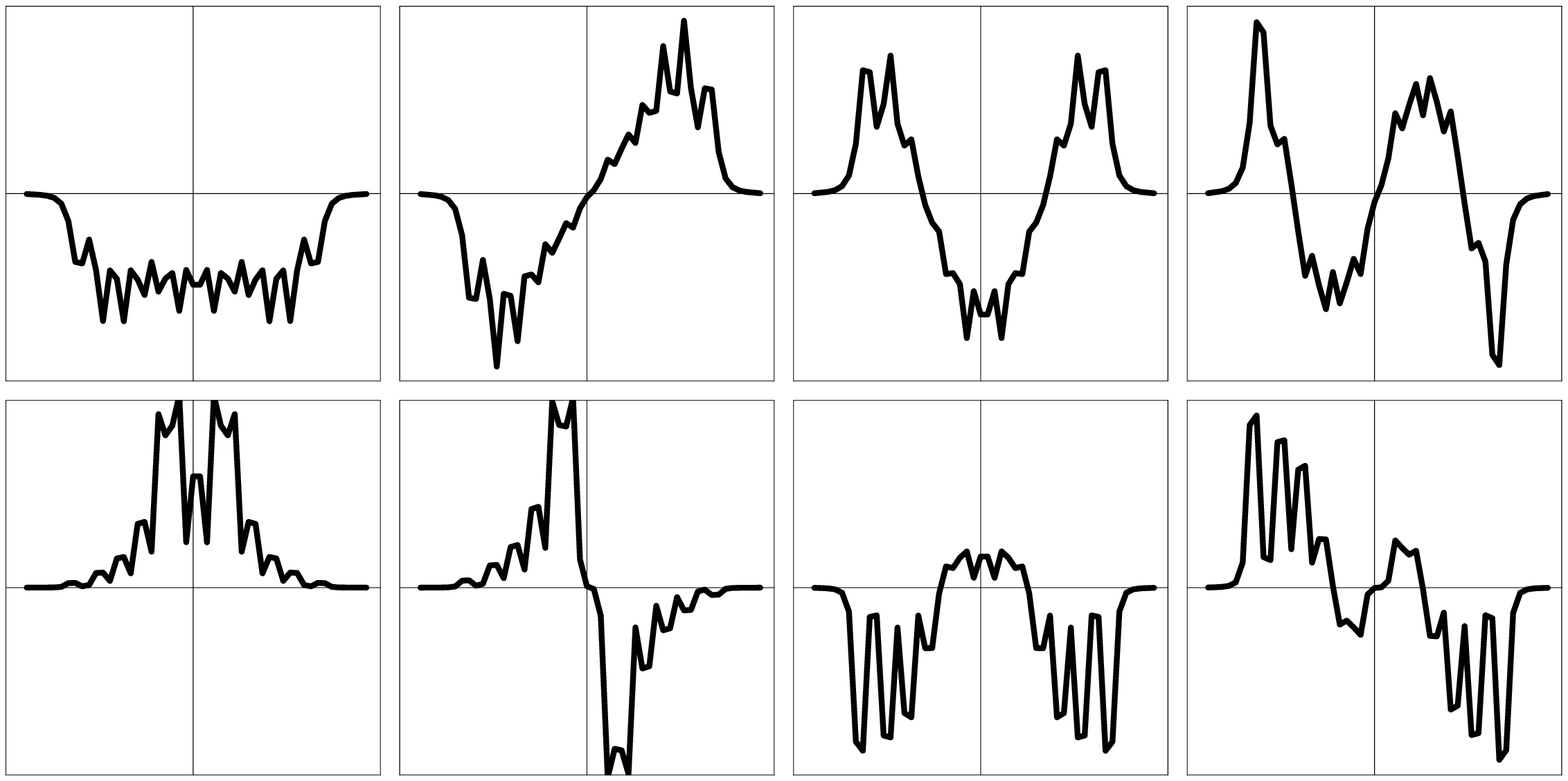}
\vglue -0.0cm
\caption{Phonon eigenmodes in the chain with
$N=50$ ions at the golden mean ion density
with $\omega=0.014,  \omega^2=1.986 \times 10^{-4}$.
Panels show the amplitude of eigenmode (in arbitrary linear units)
vs. ion positions $x_i$ varied from minimal to maximal value
for lowest 4 modes with number 0, 1, 2, 3 (from left to right).
Top row is for $K=0.03$ and bottom row is for $K=0.2$.
}
\label{fig9}
\end{figure}

\begin{figure}[t!]
\epsfxsize=3.2in
\epsffile{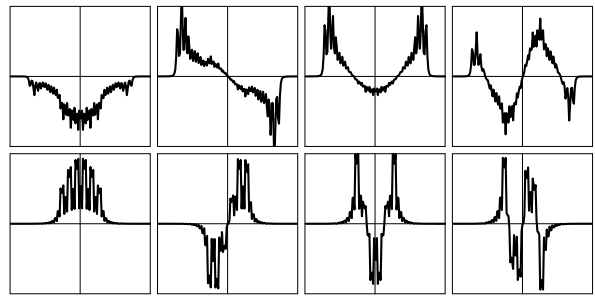}
\vglue -0.0cm
\caption{Same as in Fig.\ref{fig9} for $N=150$,
$\omega=0.00528,  \omega^2=2.794 \times 10^{-5}$.
}
\label{fig10}
\end{figure}

\begin{figure}[t!]
\epsfxsize=3.2in
\epsffile{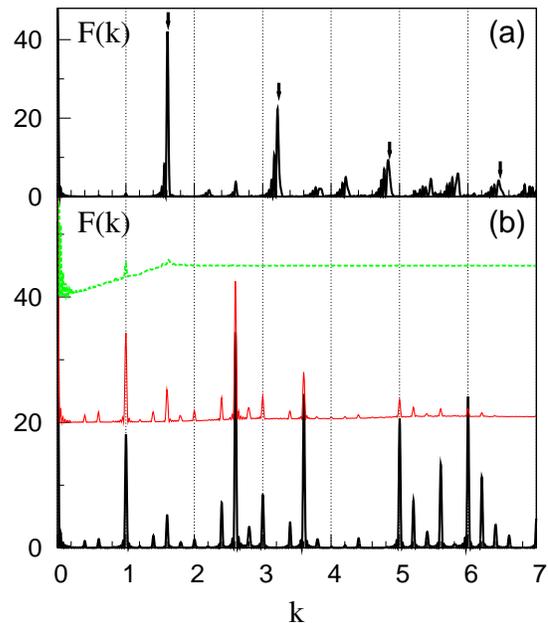}
\vglue -0.0cm
\caption{
(Color online) Formfactor $F(k)$ (see text) of the chain  with
$N=150$ ions and $\omega=0.00528, \omega^2=2.794 \times 10^{-5}$.
\textbf{\textsf{(a)}}: The classical incommensurate phase at $K=0.03$, $\hbar=0$, 
arrows mark the peaks at integer multiples of golden mean density $\nu_g$.
\textbf{\textsf{(b)}} The  pinned phase  at $K=0.2$ 
for $\hbar=0$ (bottom black curve), $\hbar=0.1$ (middle red curve
shifted 20 units upward), $\hbar=2$ (top green curve shifted 40 units upward,
for clarity $F(k)$ is multiplied by factor 5).
The temperature of the quantum chain
is $T=\hbar/\tau_0$ with $\tau_0=400$ so that $T \ll K$ and $T \ll \hbar \omega_0(K)$.
}
\label{fig11}
\end{figure}

Here in Figs.\ref{fig6}-\ref{fig11} we present additional data showing results
with larger number of ions.

%\newpage

%\end{widetext}

%\newpage


\begin{thebibliography}{99}
\bibitem{walther1992} G.~Birkl, S.~Kassner and H.~Walther, Nature {\bf 357}, 310 (1992).
\bibitem{cz1995} J.I.~Cirac and P.~Zoller, Phys. Rev. Lett. {\bf 74}, 4091 (1995).
\bibitem{blatt} H.~H\"affner, W.~H\"ansel, C.F.~Roos, J.~Benhelm, D.~Chek-al-kar,
         M.~Chwalla, T.K\"orber, U.D.~Rapol, M.~Riebe, P.O.~Schmidt, C.~Becher, O.~G\"hne, 
         W.~D\"ur and R.~Blatt, Nature {\bf 438}, 643 (2005).
\bibitem{wineland} D.~Leibfried, E.~Knill, S.~Seidelin, J.~Britton, R.B.~Blakestad,
         J.~Chiaverini, D.B.~Hume, W.M.~Itano, J.D.~Jost,
         C.~Langer, R.~Ozeri, R.~Reichle and D.J.~Wineland, Nature {\bf 438}, 639 (2005).
\bibitem{dubin} D.H.E.~Dubin and T.M.~O'Neil, Rev. Mod. Phys. {\bf 71}, 87 (1999).
\bibitem{morigi} G.~Morigi and S.~Fishman, Phys. Rev. Lett. {\bf 93}, 170602 (2004). 
\bibitem{bloch} I.~Bloch, T.W.~H\"ansch and T.~Esslinger, Nature {\bf 403}, 166 (2000).
\bibitem{walther1997} H.~Katori, S.Schlipf and H.~Walther, Phys. Rev. Lett. 
        {\bf 79}, 2221 (1997).
\bibitem{fk1938} Ya.I.~Frenkel and T.A.~Kontorova, Zh. Eksp. Teor. Fiz. {\bf 8}, 1340 (1938)
         [Phys. Z. Sowjetunion {\bf 13}, 1 (1938)].
\bibitem{braun2004} O.M.~Braun and Yu.S.~Kivshar, {\it The Frenkel-Kontorova Model: 
         Concepts, Methods, and Applications},  Springer-Verlag, Berlin (2004).
\bibitem{aubry} S.~Aubry, Physica D {\bf 7}, 240 (1983).
\bibitem{zhirov2002} O.V.~Zhirov, G.~Casati and D.L.~Shepelyansky,
         Phys. Rev. E {\bf 65}, 026220 (2002).
\bibitem{borgonovi} F.~Borgonovi, I.~Guarneri and D.L.~Shepelyansky,
         Phys. Rev. Lett. {\bf 63}, 2010 (1989).
\bibitem{berman} G.P.~Berman, E.N.~Bulgakov and D.K.~Cambell, 
        Phys. Rev. B {\bf 49}, 8212 (1994).
\bibitem{bambihu} B.~Hu, B.~Li and W.M.~Zhang, Phys. Rev. E {\bf 58}, R4068 (1998).
\bibitem{zhirov2003} O.V.~Zhirov, G.~Casati and D.L.~Shepelyansky,
         Phys. Rev. E {\bf 67}, 056209 (2003).
\bibitem{chirikov1979} B.V.~Chirikov, Phys. Rep. {\bf 52}, 263 (1979).
\bibitem{creutz} M.~Creutz and B.~Freedman, Ann. Phys. (N.Y.) {\bf 132}, 427 (1981).
\bibitem{novikov} D.S.~Novikov, Phys. Rev. Lett. {\bf 95}, 066401 (2005). 
\end{thebibliography}
\end{document}